\documentclass[
    ,final            % use final for the camera ready runs
  ]
  {aipproc}

\layoutstyle{6x9}

\def\NO{{\tilde\chi^0_1}}
\def\NT{{\tilde\chi^0_2}}

\def\lR{{\tilde l_R} }
\def\qL{{\tilde q_L}}

\def\lN{{l_{\rm n}}}
\def\lF{{l_{\rm f}}}

\def\low{{\rm low}}
\def\high{{\rm high}}

\def\mll{m_{ll}}
\def\mqll{m_{qll}}

\def\mqlLow{m_{ql(\low)}}
\def\mqlHigh{m_{ql(\high)}}

\def\max{{\rm max}}
\def\maxmll{m^\max_{ll}}
\def\maxmqll{m^\max_{qll}}
\def\maxmqlLow{m^\max_{ql(\low)}}
\def\maxmqlHigh{m^\max_{ql(\high)}}

\def\mNO{{m_\NO}}
\def\mlR{{m_\lR}}
\def\mNT{{m_\NT}}
\def\mqL{{m_\qL}}

\def\sNO{{m_\NO^2}}
\def\slR{{m_\lR^2}}
\def\sNT{{m_\NT^2}}
\def\sqL{{m_\qL^2}}

\begin{document}

\title{Mass Determination in Cascade Decays Using Shape Formulas}

\classification{11.80.Cr, 12.60.Jv, 14.80.Ly}
%\texttt{http://www.aip..org/pacs/index.html}

\keywords      {BSM, SUSY, MSSM, MSUGRA}

\author{B. K. Gjelsten}{
address={Laboratory for High Energy Physics, University of Bern, CH-3012 Bern, Switzerland}
}

\author{D. J. Miller}{
address={Department of Physics and Astronomy, University of Glasgow, Glasgow G12 8QQ, U.K.}
}

\author{P. Osland}{
address={Department of Physics and Technology, University of Bergen, N-5007 Bergen, Norway}
}

\author{A. R. Raklev}{
address={Theory Division, Physics Department, CERN, CH-1211 Gen\` eve, Switzerland}
}

\begin{abstract}
In SUSY scenarios with invisible LSP, 
sparticle masses can be determined 
from fits to the endpoints of invariant mass distributions. 
Here we discuss possible improvements by using the shapes of the distributions. 
Positive results are found 
for multiple-minima situations 
and for mass regions where the endpoints do not contain sufficient information to obtain the masses. 
\end{abstract}

\maketitle

%%%%%%%%%%%%%%%%%%%%%%%%%%%%%%%%%%%%%%%%%%%%
%% MAINMATTER
%%%%%%%%%%%%%%%%%%%%%%%%%%%%%%%%%%%%%%%%%%%%

\section{Introduction}

In R-parity conserving supersymmetric models 
sparticles are produced in pairs at the collision point, 
then decay in cascades, 
resulting in a number of Standard Model particles 
as well as two Lightest Supersymmetric Particles (LSPs), 
one for each primary sparticle. 
If the LSPs leave the detector without a trace, 
as is the case in most scenarios of this type, 
the event cannot be fully reconstructed. 
This in turn prevents a direct measurement of the sparticle masses 
from mass peaks. 
The foreseen way to obtain these masses in such scenarios, 
is through endpoint measurements of 
invariant mass distributions \cite{Hinchliffe:1996iu,Bachacou:1999zb,Allanach:2000kt,Lester:2001zx,Gjelsten:2004ki,Weiglein:2004hn,Gjelsten:2005aw}. 
Particularly suited is the cascade decay 
$\qL\to\NT q\to\lR\lN q\to\NO\lF\lN q$ 
which for a large share of SUSY models and parameters 
is kinematically allowed and appears at a rate sufficient for study. 
Four invariant mass distributions 
%\footnote{Since the two leptons can not easily be distinguished, the distributions $\mqlN$ and $\mqlF$ are inaccessible. Their replacements $\mqlHigh$ and $\mqlLow$ are built from the larger and smaller of the two $\mql$ values on an event by event basis.} 
can be constructed from the visible decay products, $\mll$, $\mqll$, $\mqlHigh$ and $\mqlLow$\footnote{Additional {\em constrained} mass distributions can be constructed, 
e.g. $\mqll$ for events with $\mll>\maxmll/\sqrt{2}$. 
The corresponding endpoint measurements however usually involve large errors. 
Although relevant, such distributions have not been considered in this study.
}. 

Once the endpoints have been measured, the masses can be found. 
However, measuring an endpoint is not trivial. 
A fit must be made to the edge region of the distribution. 
Such a fit necessarily involves a signal and a background hypothesis.  
If these are not in good correspondence with the true shapes, 
it is expected that the endpoint fitting may introduce 
large systematic uncertainties. 
While a straight-line hypothesis sometimes performs well, 
the fact that each distribution 
can take on a large variety of shapes \cite{Gjelsten:2004ki,Gjelsten:2005aw}
really is a limiting factor, 
especially since a mismeasurement of an endpoint 
by 10~GeV can easily multiply to 40~GeV in the resulting masses. 
Without an appropriate signal hypothesis the endpoint method is incomplete. 
The many studies undertaken so far reflect this 
in that they have mainly 
focused on determining the statistical error, 
which can be fairly well estimated by a straight line, 
putting off the systematics of the fitting procedure for later investigations.  
With the recent advent of an analytic description of the shapes 
of these mass distributions \cite{Miller:2005zp}, 
the time for such investigations has come closer.

\section{Masses from shapes}

The completion of the endpoint method requires formulas for the 
edge-region part of the distributions expressed by the endpoints. 
The shape formulas are however written in terms of the sparticle masses. 
Fitting the distributions gives the masses directly 
without any recourse to the endpoints. 
The considerable difference in approach 
asks for a comparison between the two methods. 
We here investigate the matter for the 
mSUGRA point SPS~1a \cite{Allanach:2002nj}. 
In order for the characteristics of the methods to easily shine through, 
we make a simple comparison, with no background included, and no detector effects. 
The comparison is based on 
histograms generated randomly from the shape formulas taken from 
\cite{Miller:2005zp} 
convoluted with a gaussian 
of width 2~GeV for $\mll$ and 10~GeV for the other distributions. 
The selected procedure and numbers mimic 
the effect of sparticle widths, initial and final-state radiation, 
detector effects etc. 
The same procedure is also invoked for the shape-fitting process.  
The normalisation of the histograms corresponds roughly to the 
ultimate statistics expected for an LHC experiment.

 \begin{table}[h]
 \begin{tabular}{lcccc}
 \hline
   & \tablehead{1}{r}{b}{Nominal}
   & \tablehead{1}{r}{b}{Fitted}
   & \tablehead{1}{r}{b}{Error}
   & \tablehead{1}{r}{b}{Inflated error}   \\
 \hline
 $\maxmll$     & 77.07 & 77.13 & 0.04    & 0.04\\
 $\maxmqlHigh$ & 375.8 & 378.9 & 0.4  & 1.2\\
 $\maxmqlLow$  & 298.5 & 304.2 & 0.8  & 2.4\\
 $\maxmqll$    & 425.9 & 432.3 & 0.3  & 0.9\\
 \hline
 \end{tabular}
 \caption{Endpoint values and errors [in GeV] for SPS~1a.}
 \label{table:endpoint}
 \end{table}

\paragraph{Endpoint method}
For the endpoint analysis a straight-line fit is used to obtain 
the statistical uncertainty. 
Table~\ref{table:endpoint} shows the nominal values 
together with the fitted values and the errors obtained. 
A second set of errors, three times the statistical ones, except for $\maxmll$, is included to give a slightly more realistic situation. 
The discrepancy between the nominal and the fitted endpoints reminds us of the problem with endpoint fitting. 
%Energy scale uncertainty is not taken into account. 
%For comparison with a more realistic endpoint analysis, see \cite{Gjelsten:2004ki}.

\begin{figure}[h]
  \includegraphics[height=.2\textheight]{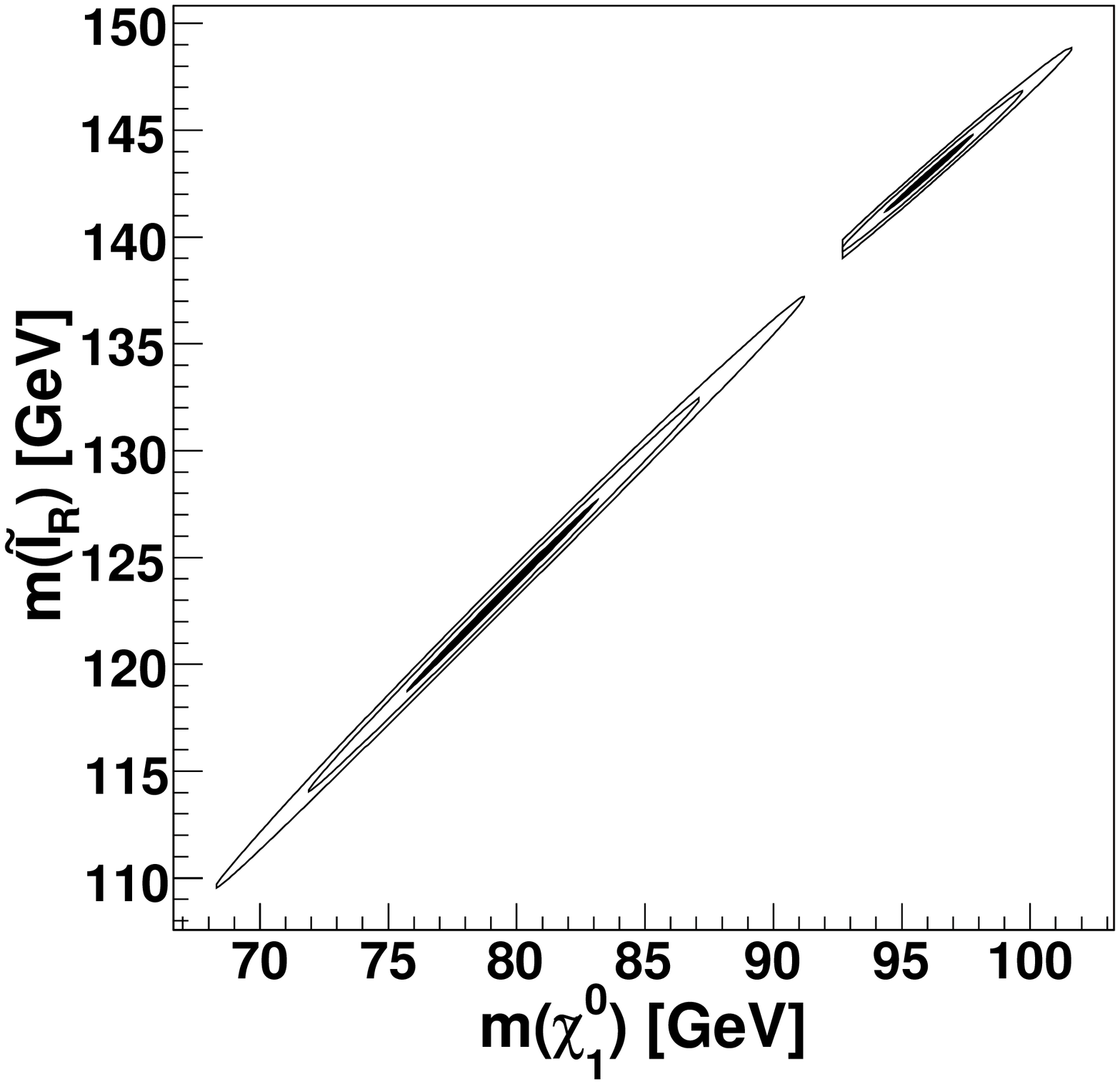}
  \includegraphics[height=.2\textheight]{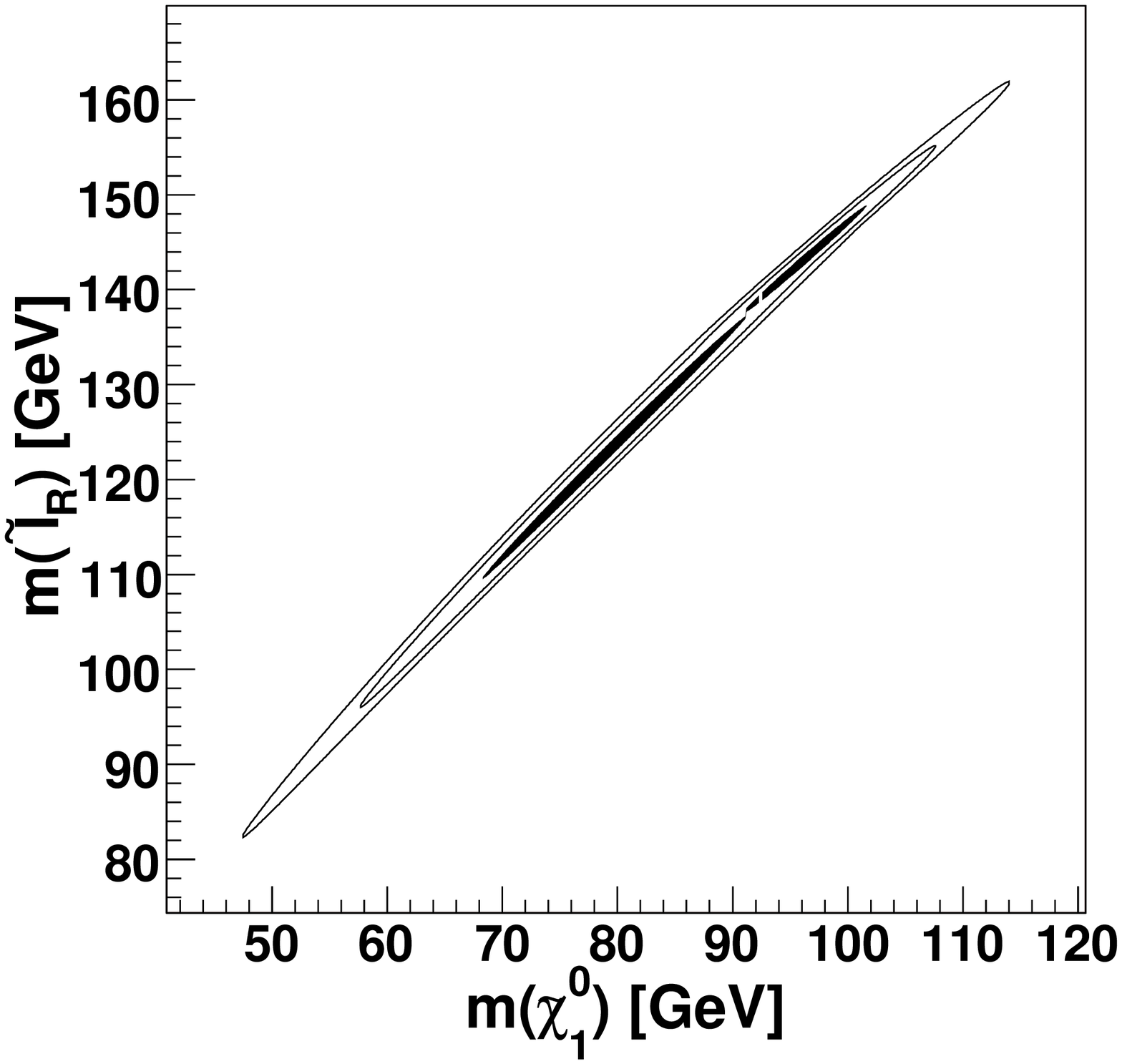}
  \includegraphics[height=.2\textheight]{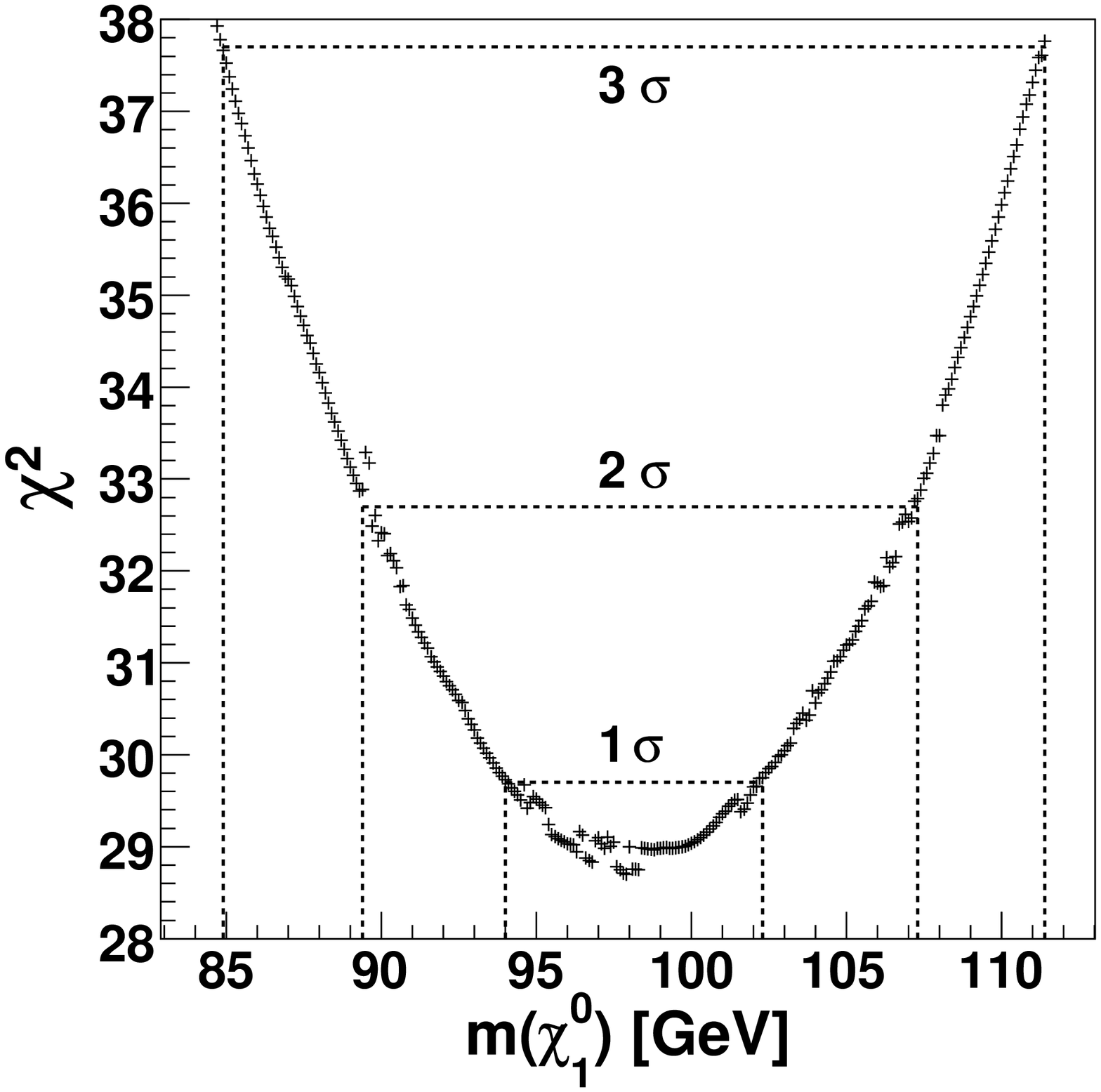}
  \caption{SPS~1a comparison. Left (Middle): endpoint method for the first (second) set or errors. Two minima are found, the correct one situated at $\mNO=96$~GeV, the false at $\mNO=80$~GeV. Right: shape fit to the same distributions. The false minimum is not present.}
  \label{fig:ellipses_endpoint}
\end{figure}

Figure~\ref{fig:ellipses_endpoint} (left and middle) shows the resulting 
1, 2 and 3$\sigma$ 
contours in the $\mNO$--$\mlR$ plane for the two sets of errors. 
The nominal endpoint values are used, not the fitted ones. 
From these figures the main characteristics of the endpoint method are apparent: 
the occurrence of 
multiple solutions \cite{Gjelsten:2004ki,Gjelsten:2005sv}
and strong mass correlation 
reflecting the fact that 
mass differences are accurately determined 
while the overall scale only poorly so. 

\paragraph{Shape method}
Since the background is expected to be more prominent for lower invariant masses, 
only the higher part of the distributions are used in the shape fit. 
The result of the fit is given by the $\chi^2$-function in 
Fig.~\ref{fig:ellipses_endpoint} (right). 
Notice that the false solution is absent. 
The shape difference for the two mass sets is apparently sufficient 
to discriminate between them. 
Whether or not this situation holds for a more realistic analysis 
remains to be seen. The shape difference between these two points is 
below mSUGRA average \cite{Raklev2006}. 
The precision on the overall mass scale 
returned by the shape fit 
is roughly as for the endpoint approach using the second error set, 
except for the false solution present in the endpoint case 
which stretches to very low masses.
Finally it is found that also for the shape fitting method 
the resulting masses are very correlated, 
constraining mass differences much more than masses. 
More study is however needed to compare the 
degree of 
correlation to that obtained from the endpoint method. 

\section{When endpoints are not enough}
In some regions of mass space 
the four endpoints are no longer linearly independent 
due to $(\maxmqll)^2 = (\maxmll)^2 + (\maxmqlHigh)^2$. 
When this happens, which is for regions {\it(2,3)}, {\it(3,1)} and {\it(3,2)} 
in the notation of \cite{Gjelsten:2004ki}, 
curves exist in the four-dimensional mass space 
on which all mass points produce the same endpoints. 
In region {\it(3,2)} the curve is given by 
\begin{equation}
\def\mquad{\phantom{ee}}
\slR=\frac{\sNO}{2-A},\mquad
\sNT=\slR + \frac{(\maxmll)^2}{A-1},\mquad 
\sqL=\slR + \frac{(\maxmqll)^2}{A-1},\mquad
A=\left(\frac{\maxmqlHigh}{\maxmqlLow}\right)^2
\label{eq:region32}
\end{equation}
From one mass set new mass sets on the curve can be generated by keeping the endpoints fixed, then changing $\mNO$ and calculating the other masses. 
\begin{figure}
  \includegraphics[height=.16\textheight] {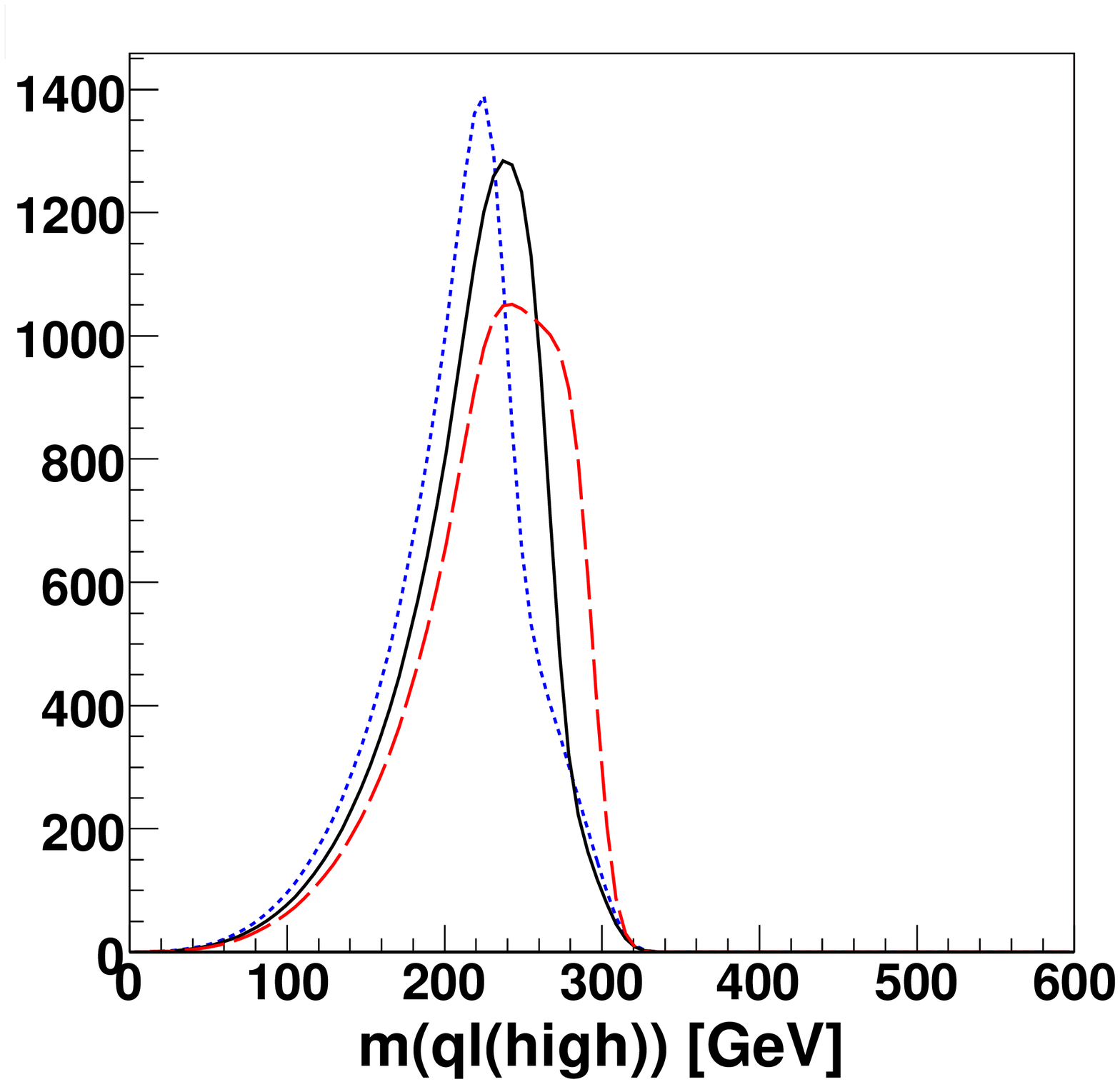}
  \includegraphics[height=.16\textheight] {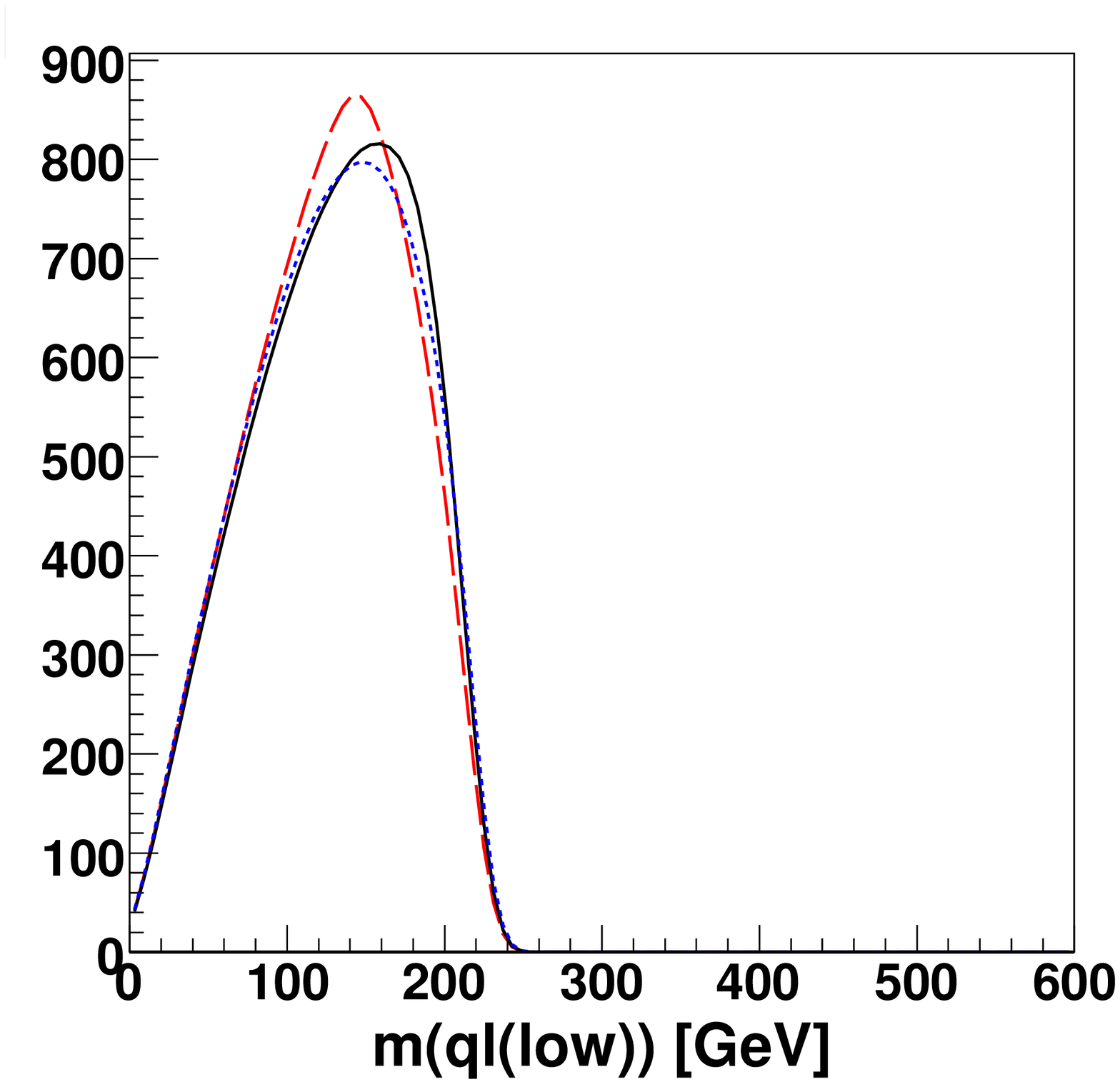}
  \includegraphics[height=.16\textheight] {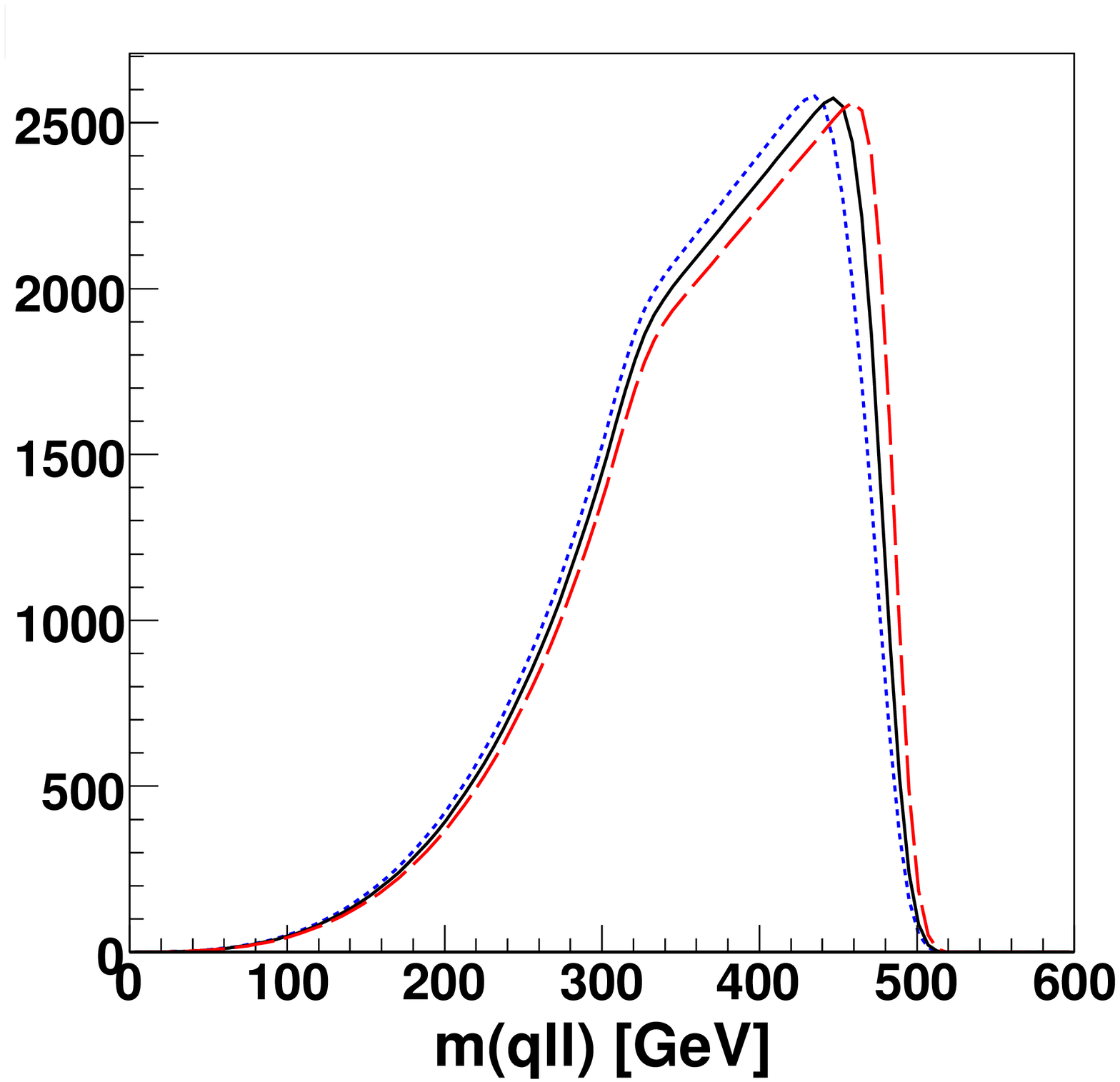}
  \includegraphics[height=.16\textheight] {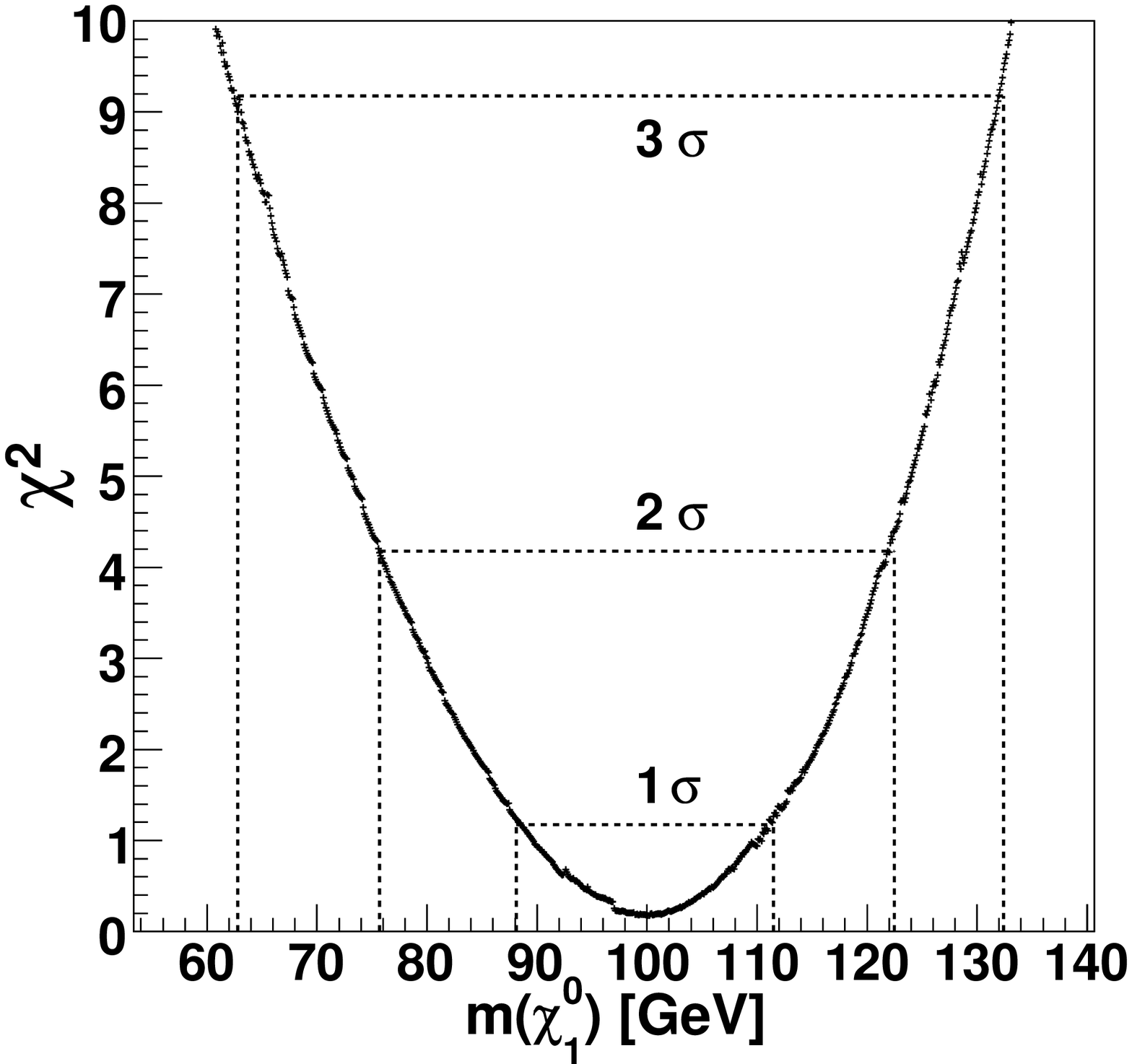}
  \caption{Left to right: $\mqlHigh$, $\mqlLow$ and $\mqll$ for three points 
in region {\it(3,2)}, all having the same endpoints, but different distributions. Right: $\chi^2$-function from fitting the shapes has minimum at the correct  $\mNO$. Only smearing is applied, no statistical fluctuations, therefore $\chi^2_{\min}=0$.}
  \label{fig:region32}
\end{figure}
Under these circumstances four endpoint measurements do not 
impose sufficient constraints 
to obtain the masses. 
Extra information is however available from the shapes. 
Figure~\ref{fig:region32} (three leftmost) shows the relevant 
gaussian-convoluted (width 10~GeV) 
mass distributions for the mass set 
$(\mNO,\mlR,\mNT,\mqL)=(100,300,500,600)$~GeV (black, solid), 
and two sets on the line~(\ref{eq:region32}) 
defined by increasing (blue, dotted) and decreasing (red, dashed) 
$\mNO$ by 30~GeV. 
While the endpoints are the same, the shapes are seen to differ. 
(The $\mll$ distribution is not shown as its shape is always the same.) 

%%%The shape difference of the $\mqlHigh$ distribution is the largest. 
Based on the distributions of the original mass set ($\mNO=100$~GeV, black solid), 
histograms are generated, assuming a similar number of events as for the 
previous SPS~1a investigations. 
Figure~\ref{fig:region32} (right) shows the $\chi^2$ function from a shape fit 
to these distributions. 
The histograms contain no statistical fluctuations, 
which is why the minimum is at $\chi^2=0$. 
Adding statistical fluctuations will lift $\chi^2_{\min}$ from zero, 
but the conclusion stays the same: 
A clear minimum is found for the correct masses 
in this region where the endpoints alone can not bring the masses.

\section{Conclusion}
We have investigated the use of shape formulas 
to obtain the sparticle masses in scenarios where the LSP is undetected. 
Comparison with the standard endpoint method was made for the 
mSUGRA point SPS~1a. 
The extra solution at $\mNO=80$~GeV 
which is returned by the endpoint inversion, 
is absent in the shape-fitting results. 
The strong correlation in the resulting masses remains: 
Mass differences are accurately determined 
while the overall mass scale has larger uncertainties. 
It was furthermore shown that the use of shapes allows for 
masses to be determined in regions of mass space where the 
endpoints alone do not suffice due to linear interdependence. 
In conclusion, the use of shapes to determine masses 
constitute a very promising approach. 
More study is needed. 
It is particularly important to understand what impact 
the distortion of the invariant mass distributions, 
particularly from selection cuts and various backgrounds, 
will have on the proposed shape fitting technique.

%%%%%%%%%%%%%%%%%%%%%%%%%%%%%%%%%%%%%%%%%%%%%%%%
%% BACKMATTER
%%%%%%%%%%%%%%%%%%%%%%%%%%%%%%%%%%%%%%%%%%%%%%%%

\begin{theacknowledgments}
BKG would like to thank Bern LHEP colleagues for useful discussions. 
ARR acknowledges support from the European Community through a Marie
Curie Fellowship for Early Stage Research Training.
This research has been supported in part by the Research Council of Norway 
and the Swiss National Science Foundation. 
\end{theacknowledgments}


\end{document}